\documentclass[twocolumn,showpacs,aps,prl,superscriptaddress]{revtex4}

\usepackage{graphicx}
\usepackage{dcolumn}
\usepackage{amsmath}
\usepackage{epsfig}

\input pubboard/babarsym
\newcommand{\beq}{\begin{equation}}
\newcommand{\eeq}{\end{equation}}
\newcommand{\bea}{\begin{eqnarray}}
\newcommand{\eea}{\end{eqnarray}}

\newcommand{\DE}{\ensuremath{\Delta E}}

\newcommand{\pvec}{{\bf p}}

\def\Y#1S{{\Upsilon\rm(#1S)}}

\newcommand{\BABARPubYear}    {07}
\newcommand{\BABARPubNumber}  {055}

\newcommand{\SLACPubNumber} {11965}

\def\figurebox#1#2#3{%
    \def\arg{#3}%
    \ifx\arg\empty
    {\hfill\vbox{\hsize#2\hrule\hbox to #2{\vrule\hfill\vbox to #1{\hsize#2\vfill}\vrule}\hrule}\hfill}%
    \else
    {\hfill\epsfbox{#3}\hfill}%
    \fi}

\begin{document}

\preprint{\babar-PUB-\BABARPubYear/\BABARPubNumber} 
\preprint{SLAC-PUB-\SLACPubNumber} 

\noindent \babar-PUB-07/055\\ 
SLAC-PUB-12746\\

\title{
{\large \bf \boldmath
Measurements of the Branching Fractions of \\ \Bz \to $K^{*0}K^+K^-$, \Bz \to $K^{*0}\pi^+K^-$, \Bz \to $K^{*0}K^+\pi^-$, and \Bz \to $K^{*0}\pi^+\pi^-$} 
}

%
\author{B.~Aubert}
\author{M.~Bona}
\author{D.~Boutigny}
\author{Y.~Karyotakis}
\author{J.~P.~Lees}
\author{V.~Poireau}
\author{X.~Prudent}
\author{V.~Tisserand}
\author{A.~Zghiche}
\affiliation{Laboratoire de Physique des Particules, IN2P3/CNRS et Universit\'e de Savoie, F-74941 Annecy-Le-Vieux, France }
\author{J.~Garra~Tico}
\author{E.~Grauges}
\affiliation{Universitat de Barcelona, Facultat de Fisica, Departament ECM, E-08028 Barcelona, Spain }
% ...
\collaboration{The \babar\ Collaboration}
\noaffiliation

\date{\today}

\begin{abstract}
Branching fraction measurements of charmless $B^0\rightarrow K^{*0}h^+_1h^-_2$ ($h_{1,2}$ = $K$, $\pi$) decays are presented, using a data sample of 383 million $\Upsilon(4S) \rightarrow$ \BB\ decays collected with the \babar\ detector at the PEP-II asymmetric-energy $B$-meson factory at SLAC. The results are: ${\cal B}$($B^0 \rightarrow K^{*0}K^+ K^-)$ = (27.5 $\pm$ 1.3 $\pm$ 2.2) $\times$ 10$^{-6}$,  ${\cal B}$($B^0$ $\rightarrow$ $K^{*0}\pi^+ K^-$) = (4.6 $\pm$ 1.1 $\pm$ 0.8) $\times$ 10$^{-6}$ and ${\cal B}$($B^0$ $\rightarrow$ $K^{*0}\pi^+\pi^-$) = (54.5 $\pm$ 2.9 $\pm$ 4.3) $\times$ 10$^{-6}$. The first errors quoted are statistical and the second are systematic. An upper limit is set for ${\cal B}$($B^0$ $\rightarrow$ $K^{*0}K^+ \pi^-$) $<$ 2.2 $\times$ 10$^{-6}$ at 90\% confidence level. We also present measurements of $C\!P$-violating asymmetries for the observed decays.
\end{abstract}


\maketitle
\begin{figure}[htb]
\resizebox{\columnwidth}{!}{
\includegraphics{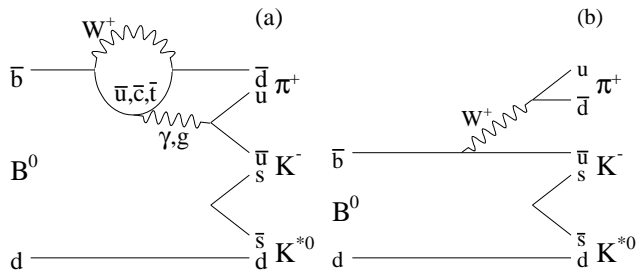}
}
\caption{(a) Penguin and (b) tree Feynman diagrams for the decay $B^0$ $\rightarrow$ $K^{*0}\pi^+ K^-$. Similar Standard Model diagrams do not exist for the $\Delta S$ = 2 decay $B^0$ $\rightarrow$ $K^{*0}K^+ \pi^-$.
}\label{fig:feynKstkpi}
\end{figure}

 Charmless decays of $B$ mesons to three-body final states are important probes of the weak interaction and the complex quark couplings of the Cabibbo--Kobayashi--Maskawa (CKM) matrix~\cite{ckm}. Improved experimental measurements of these charmless decays, combined with theoretical developments, can provide significant constraints on the CKM matrix elements and potentially uncover evidence for physics beyond the Standard Model. For example, the branching fraction of the decay $B^0$ $\rightarrow$ $K^{*0}\pi^+ K^-$ is sensitive to the CKM matrix elements $V_{td}$ and $V_{ub}$ (see Fig.~\ref{fig:feynKstkpi}). Additionally, a branching fraction of the Standard Model suppressed decay $B^0$ $\rightarrow$ $K^{*0}K^+ \pi^-$ comparable or larger than that of $B^0$ $\rightarrow$ $K^{*0}\pi^+ K^-$ would be an indication of new physics.

Neutral $B$-meson decays to $K^+\pi^-h^+_1 h^-_2$ (where $h_{1,2}$ = $K$ or $\pi$) are dominated by $K^{*0}h^+_1 h^-_2$, but can also proceed via a four-body nonresonant component, through intermediate charmless resonances such as $B^0 \rightarrow K^{*0}\phi$ or $B^0 \rightarrow K^{*0}\rho^0$~\cite{babarkstphi,babarkstrho,bellekstphi}, or other as-yet-unobserved intermediate charmless resonances. To date, there are only limits on the inclusive decays $B^0 \rightarrow K^{*0}h^+_1 h^-_2$, measured by the ARGUS experiment using less than 0.2 fb$^{-1}$~\cite{argus1991}, whilst there are more accurate measurements of charged $B$-meson decays to $K^{*+}h^+_1 h^-_2$~\cite{kstarp}.

We identify $K^{*0}$ mesons through their decay to $K^+\pi^-$; consequently, the states $B^0\rightarrow K^{*0}h^+_1h^-_2$ are self-tagged as the charge of the daughter kaon from the $K^*$ reflects the flavor of the \Bz meson. The $C\!P$-violating decay rate asymmetry is defined as

\begin{equation}
{\cal A}_{\Kstar h_1h_2} \equiv \frac{\Gamma_{\Kstarzb h^-_1h^+_2} - \Gamma_{K^{*0}h^+_1h^-_2}}{\Gamma_{\Kstarzb h^-_1h^+_2} + \Gamma_{K^{*0}h^+_1h^-_2}} ~,
\end{equation}

\noindent where $\Gamma$ is the partial $B$ decay width. 

In this paper, branching fractions of $B^0 \rightarrow K^{*0}K^+ K^-$, $B^0 \rightarrow K^{*0}\pi^+ K^-$ and $B^0 \rightarrow K^{*0}\pi^+ \pi^-$ are measured, and an upper limit is set for the Standard Model suppressed decay $B^0 \rightarrow K^{*0}K^+ \pi^-$~\cite{cc}. The selection criteria require events with a reconstructed $K^+\pi^- h^+_1 h^-_2$ final state, so that the total charmless contribution (including resonant charmless substructure) to the $K^{*0}h^+_1h^-_2$ Dalitz plot can be measured. The ${\cal A}_{K^*h_1h_2}$ values for the observed decays $B^0 \rightarrow K^{*0}K^+ K^-$, $B^0 \rightarrow K^{*0}\pi^+ K^-$, and $B^0 \rightarrow K^{*0}\pi^+ \pi^-$ are also measured.

The data on which this analysis is based were collected with the \babar\ detector~\cite{babar} at the \pep2\ asymmetric-energy \epem\ storage ring. The \babar\ detector consists of a double-sided five-layer silicon tracker, a 40-layer drift chamber, a Cherenkov detector, an electromagnetic calorimeter, and a magnet with instrumented flux return (IFR) consisting of layers of iron interspered with resistive plate chambers or limited streamer tubes. The data sample has an integrated luminosity of 348~fb$^{-1}$ collected at the $\FourS$ resonance, which corresponds to $(383 \pm 4 )\times 10^6$ \BB\ pairs. It is assumed that the $\FourS$ decays equally to neutral and charged $B$-meson pairs. In addition, 37 fb$^{-1}$ of data collected 40~MeV below the $\FourS$ resonance are used for background studies.

Candidate $B$ mesons are reconstructed from four tracks that are consistent with originating from a common decay point within the PEP-II luminous region. Each of the four tracks is required to have at least 12 hits in the drift chamber and a transverse momentum greater than 100~\mevc.  The tracks are identified as either pion or kaon candidates, with protons vetoed, using Cherenkov-angle information and ionization energy-loss rate measurements (\dedx). The efficiency for kaon selection is approximately 80\%, including geometric acceptance, while the probability of misidentification of pions as kaons is below 5\% up to a laboratory momentum of 4\gevc. Muons are rejected using information predominantly from the IFR. Furthermore, the tracks are required to fail an electron selection based on their ratio of energy in the calorimeter to momentum in the drift chamber, shower shape in the calorimeter, \dedx, and Cherenkov-angle information. 

To characterize signal events, three kinematic variables and one event-shape variable are used. The first kinematic variable, $\DeltaE$, is the difference between the center-of-mass (c.m.)\ energy of the $B$ candidate and $\sqrt{s}/2$, where $\sqrt{s}$ is the total c.m.\ energy. The second is the beam-energy-substituted mass $\mes = \sqrt{(s/2 + \pvec_i \cdot \pvec_B)^2/E_i^2 - \pvec^2_B}$, where  $\pvec_B$ is the $B$ momentum and  ($E_i, \pvec_i$) is the four-momentum of the $\FourS$ in the laboratory frame.  The third kinematic variable is the $K^+ \pi^-$ invariant mass, $m_{\Kstarz}$, used to identify \Kstarz candidates. $B$ Candidates are required to be in the ranges  $|\Delta E| <0.1 \gev$, $5.2500<\mes<5.2889 \gevcc$, and 0.776 $<$ $m_{\Kstarz}$ $<$ 0.996 GeV/$c^2$. The event-shape variable is a Fisher discriminant $\mathcal{F}$~\cite{Fisher}, constructed as a linear combination of the absolute value of the cosine of the angle between the $B$ candidate momentum and the beam axis, the absolute value of the cosine of the angle between the thrust axis of the decay products of the $B$ candidate and the beam axis, and the zeroth and second angular moments of energy flow about the thrust axis of the reconstructed $B$.

Continuum quark production ($e^+e^-$ $\rightarrow$ $q\bar{q}$, where $q$ = {\em u,d,s,c}) is the dominant source of background. It is suppressed using another event-shape variable, $|\cos\theta_T|$, which is the absolute value of the cosine of the angle $\theta_T$ between the thrust axis of the selected $B$ candidate and the thrust axis of the rest of the event. For continuum background, the distribution of $|\cos\theta_T|$ is strongly peaked towards 1.0 whereas the distribution is essentially flat for signal events. Therefore, the relative amount of continuum background is reduced by requiring $|\cos\theta_T| < 0.8$.

Monte Carlo (MC) events that are at least a 1000 times the number expected in data are used to study background from other $B$-meson decays. The largest $B$-background for $B^0 \rightarrow K^{*0}\pi^+ \pi^-$ candidates comes from decays including charmonium mesons, such as $J/\psi K^{*0}$, $\chiczero K^{*0}$ and $\psi(2S)K^{*0}$, where charmonium decays to $\mu^+\mu^-$ are misidentified as decays to $\pi^+\pi^-$, or where the charmonium decays directly to $\pi^+\pi^-$. These background events are removed by vetoing reconstructed $\pi^+\pi^-$ masses in the ranges 3.00 $<$ $m_{\pi^+\pi^-}$ $<$ 3.20~GeV/$c^2$, 3.35 $<$ $m_{\pi^+\pi^-}$ $<$ 3.50~GeV/$c^2$ and 3.60 $<$ $m_{\pi^+\pi^-}$ $<$ 3.78~GeV/$c^2$, corresponding to the $J/\psi$, $\chiczero$ and  $\psi(2S)$ meson masses, respectively. For $B^0 \rightarrow K^{*0}K^+ K^-$ candidates, $\chiczero K^{*0}$ events are removed by rejecting events with a reconstructed invariant mass in the range 3.32 $<$ $m_{K^+K^-}$ $<$ 3.53~GeV/$c^2$. 

Potential charm contributions from $B^0$ $\rightarrow$ $D^-(\rightarrow K^{*0}h^-_2)h^+_1$ events are removed from corresponding $B^0 \rightarrow K^{*0}h^+_1 h^-_2$ candidates by vetoing events with a  reconstructed $K^{*0}h^-_2$ invariant mass in the range 1.83 $<$ $m_{\Kstar h}$ $<$ 1.91~GeV/$c^2$. To remove background from \Dz mesons, a veto is applied to any $K\pi$ pair with an invariant mass in the range 1.83 $<$ $m_{K\pi}$ $<$ 1.91~GeV/$c^2$ for each  $B^0\rightarrow K^{*0}h^+_1h^-_2$ decay. Studies of MC events show the largest remaining charmed background is  $B^0$ $\rightarrow$ $D^-(\rightarrow K^{*0}\pi^-)\pi^+$, with 11\% passing the veto. Surviving charmed events have a reconstructed $D$ mass outside the veto range as a result of using a wrong $\pi^-$ or $K^+$ candidate that is incorrectly selected from the other $B$ decay in the event. 

A fraction of events for all decay modes have more than one $\Bz$ candidate reconstructed. For those events, the candidate with the smallest $\chi^2$ of the fitted $B$ decay vertex is selected. Studies of MC events show the selection of $B^0 \rightarrow K^{*0}\pi^+ K^-$ events produce the largest number of multiple candidates, in 21\% of events, where for these multiple candidates the correct one is selected 65\% of the time.

After all requirements have been applied, there are five main sources of remaining $B$ background: two-body decays proceeding via a charmonium meson; two and three-body decays proceeding via a $D$ meson; combinatorial background from three unrelated particles ($K^{*0}h^+_1 h^-_2$);  charmless two or four-body $B$ decays with an extra or missing particle and three-body decays with one or more particles misidentified. Along with selection efficiencies obtained from MC simulation, existing branching fractions for these modes \cite{hfag,pdg} are used to estimate their background contributions that are included in fits to data. 

In order to extract the signal event yield for the channel under study, an unbinned extended maximum likelihood fit is used. The likelihood function for $N$ candidates is

\begin{equation}
  \label{eq:Likelihood}
  \mbox{$\mathcal{L}$} \,=\, \frac{1}{N!}\exp\left(-\sum_{i=1}^{M} n_i\right)\, \prod_{j=1}^N 
\,\left(\sum_{i=1}^M n_{i} \, P_{i}(\vec{\alpha},\vec{x_j})\right) ~,
\end{equation}

\noindent where $M=3$ is the number of hypotheses (signal, continuum background, and $B$-background), $n_i$ is the number of events for each hypothesis determined by maximizing the likelihood function, and $P_{i}(\vec{\alpha},\vec{x_j})$ is a probability density function (PDF) with the parameters $\vec{\alpha}$ and $\vec{x}$ = (\mes, \DE, $\mathcal{F}$ and $m_{\Kstarz}$). The PDF is a product $P_{i}(\vec{\alpha},\vec{x}) = P_{i}(\vec{\alpha}_{\mes} ,\mes ) \times P_{i}(\vec{\alpha}_{\DE} ,\DE ) \times  P_{i}(\vec{\alpha}_{\mathcal{F}}, \mathcal{F}) \times  P_{i}(\vec{\alpha}_{m_{\Kstarz}}, {m_{\Kstarz}})$. Studies of MC simulations show that correlations between these variables are small for the signal and continuum background hypotheses. However, for $B$-background, correlations are observed between \mes and \DE, which are taken into account by forming a 2-dimensional PDF for these variables. 

The parameters for signal and $B$-background PDFs are determined from MC simulation. All continuum background parameters are allowed to vary in the fit, in order to help reduce systematic effects from this dominant event type. Sideband data, defined to be in the region $0.1 < \Delta E <0.3 \gev$ and $5.25<\mes<5.29 \gevcc$, as well as data collected 40~MeV below the $\FourS$ resonance, are used to model the continuum background PDFs. For the \mes\ PDFs, a Gaussian distribution is used for signal, a threshold function~\cite{argus} for continuum and a two-dimensional histogram for $B$-background. For the \DE\ PDFs, a sum of two Gaussian distributions with distinct means is used for the signal, a first-order polynomial for the continuum background and a two-dimensional histogram is used for $B$-background. The signal, continuum and $B$-background $\cal{F}$ PDFs are described using a sum of two Gaussian distributions with distinct means and widths. Finally, for $m_{\Kstarz}$ PDFs, a sum of the Breit-Wigner function and a first-order polynomial describes the signal, continuum, and $B$-background distributions. The first-order polynomial component of the $m_{\Kstarz}$ PDFs is used to model misreconstructed events in signal and background. Within the $m_{\Kstarz}$ fit range, there is also the possibility of $B$-background contributions from nonresonant and higher $K^{*0}$ resonances, which are modeled in the fit using the LASS parameterization~\cite{aston,latham}. The contribution from this background is estimated by extrapolating a $K\pi$ invariant mass projection fitted in a higher-mass region (0.996 $<$ $m_{\Kstarz}$ $<$ 1.53 \gevcc) into the signal region. This estimated background is modeled in the final fit into the signal region, and assumes there are no interference effects between the $K\pi$ background and the $K^{*0}(892)$ signal.

Branching fractions $\mathcal{B}$ are usually calculated with the equation $\mathcal{B}$\,=\ $n_{\mathrm{sig}}/(N_{\myBB} \times \epsilon$), where $n_{\mathrm{sig}}$ is the fitted number of signal events $n_1$, $\epsilon$ is the average signal efficiency obtained from MC simulation and $N_{\myBB}$ is the total number of \BB\ events. For the $B^0 \rightarrow K^{*0}h^+_1 h^-_2$  branching fraction, the average efficiency cannot be taken directly from MC events. This is due to the efficiency variations across the Dalitz plane and because the distribution of events in the Dalitz plane is a priori unknown. To calculate the branching fraction, a weight is assigned to each event $j$ as 

\begin{equation}
\label{eq:sweight}
{\cal W}_j = \frac{\sum_iV_{1,i}P_{i}(\vec{\alpha},\vec{x_j})}{\sum_kn_{k}{P_{k}(\vec{\alpha},\vec{x_j})}} ~,
\end{equation}

\noindent where $V_{1,i}$ is the row of the covariance matrix associated with the $n_1$ parameter, obtained from the fit~\cite{splot}. This procedure is effectively a background subtraction where the weights have the property $\sum_j{\cal W}_j = n_{\mathrm{sig}}$. The branching fraction is then calculated as ${\cal B} = \sum_{j}{\cal W}_j/(\epsilon_{j} \times N_{\myBB})$,  where $\epsilon_j$ (a function of $m_{K^{*0}h^+_1}^2$, $m_{K^{*0}h^-_2}^2$ and the $K^{*0} \to K^+\pi^-$ decay helicity angle) varies across phase space and is simulated in bins using over eight million MC events for each channel. The sizes of the bins are optimized to provide continuous coverage of the efficiency distribution.

\begin{figure}[htb]
\resizebox{\columnwidth}{!}{
\includegraphics{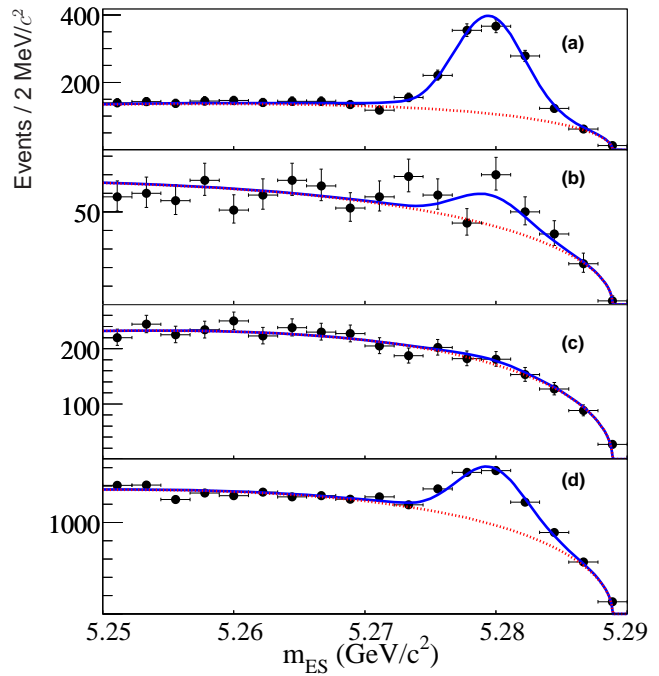}
}
\caption{Maximum likelihood fit projections of  $m_{ES}$ for signal-enhanced samples of charmless $B^0 \to K^{*0}h^+_1 h^-_2$ candidates. The dashed line is the fitted background PDF while the solid line is the sum of the signal and background PDFs. The points indicate the data. The plot shows projections for (a) $B^0 \to K^{*0}K^+ K^-$, (b) $B^0 \to K^{*0}\pi^+ K^-$, (c) $B^0 \to K^{*0}K^+ \pi^-$  and (d) $B^0 \to K^{*0}\pi^+ \pi^-$.
}\label{fig:fitproj}
\end{figure}

Figure~\ref{fig:fitproj} shows the fitted \mes projections for the $B^0 \rightarrow K^{*0}K^+ K^-$, $B^0 \rightarrow K^{*0}\pi^+ K^-$, $B^0 \rightarrow K^{*0}K^+ \pi^-$, and $B^0 \rightarrow K^{*0}\pi^+ \pi^-$  candidates, while the fitted signal yields, measured branching fractions, upper limits  and asymmetries are shown in Table \ref{tab:results}. The candidates in Fig.~\ref{fig:fitproj} are signal-enhanced, with a requirement on the probability ratio ${\cal P}_{\mathrm{sig}}/({\cal P}_{\mathrm{sig}} +{\cal P}_{\mathrm{bkg}})$, optimized to enhance the visibility of potential signal, where ${\cal P}_{\mathrm{sig}}$ and ${\cal P}_{\mathrm{bkg}}$ are the signal and the total background probabilities, respectively (computed without using the variable plotted). The 90\% confidence level (C.L.) branching fraction upper limit (${\cal B}_{\rm UL}$) is determined by integrating the likelihood distribution (with systematic uncertainties included) as a function of the branching fraction from 0 to ${\cal B}_{\rm UL}$, so that $\int^{{\cal B}_{\rm UL}}_0 {\cal L}\mathrm{d}{\cal B} = 0.9 \int^\infty_0 {\cal L}\mathrm{d}{\cal B}$. The signal significance $S$ is defined as $\sqrt{2\Delta\ln{\cal L}}$, where $\Delta\ln{\cal L}$ represents the change in log--likelihood (with systematic uncertainties included) between the maximum value and the value when the signal yield is set to zero.

\begin{table*}[!ht]
\caption{Signal yields, $B$-background yields ($B$ bkg), efficiencies, and branching fractions (\cal{$B$}), measured using $K^+\pi^- h^+_1 h^-_2$ events. Fit bias corrections are applied to the signal yields and branching fractions. The first error is statistical and  the second error is systematic. The efficiencies take into account ${\cal B}(K^{*0} \to K^+\pi^-) = 2/3$, assuming isospin symmetry. The significance $S$ is shown for $B^0 \rightarrow K^{*0}\pi^+ K^-$ and $B^0 \rightarrow K^{*0}K^+ \pi^-$ and the branching fraction upper limit at 90\% C.L. is shown for $B^0 \rightarrow K^{*0}K^+\pi^-$. Asymmetries (${\cal A}_{K^*hh}$) are reported  only for the channels with significant yields.}\label{tab:results}
\begin{center}
\begin{tabular}{ccccccccc}
\hline
\hline
Mode&
Total& 
$B$ bkg&
Signal&
Signal&
${\cal B}$ &
${\cal B}_{\rm UL}$&
$S$ ($\sigma$)&
Asymmetry (${\cal A}_{K^*hh}$)\\
&
Events&
&
Yield&
Efficiency(\%)&
($\times$ 10$^{-6}$)&
($\times$ 10$^{-6}$)&
&
\\
\hline
$B^0 \rightarrow K^{*0}K^+ K^-$&
11185& 
221&
984 $\pm$ 46&
9.3&
27.5 $\pm$ 1.3 $\pm$ 2.2&
--&
$>$ 10 &
0.01 $\pm$ 0.05 $\pm$ 0.02\\
$B^0 \rightarrow K^{*0}\pi^+ K^-$ &
53991&
276&
183 $\pm$ 42.4&
10.4&
4.6 $\pm$ 1.1 $\pm$ 0.8 &
--&
5.3&
0.22 $\pm$ 0.33 $\pm$ 0.20\\
$B^0 \rightarrow K^{*0}K^+ \pi^-$ &
16248&
372&
18.8 $\pm$ 29.4&
9.8&
0.5 $\pm$ 0.8 $\pm$ 0.5 &
2.2&
0.9&
--\\
$B^0 \rightarrow K^{*0}\pi^+ \pi^-$ &
100750&
1701&
2019 $\pm$ 108&
9.7&
54.5 $\pm$ 2.9 $\pm$ 4.3&
--&
$>$ 10&
0.07 $\pm$ 0.04 $\pm$ 0.03\\
\hline
\hline
\end{tabular}
\end{center}
\end{table*}

Contributions to the branching fraction systematic uncertainty are shown in Table~\ref{tab:sys}. Errors due to tracking efficiency are assigned by comparing control channels in MC simulation and data. The error in the efficiency is due to limited MC statistics of the signal event samples. The number of \BB\ events is calculated with an uncertainty of 1.1\%. To calculate errors due to the fit procedure, a large number of MC samples are used, containing the numbers of signal and continuum events measured in data and the estimated number of exclusive $B$-background events. The differences between the generated and fitted values are used to estimate small fit biases (see Table~\ref{tab:sys}) that are a consequence of correlations between fit variables. These biases are applied as corrections to obtain the final signal yields, and half of the correction is added as a systematic uncertainty. The uncertainty of the $B$-background contribution to the fit is estimated by varying the known branching fractions within their errors. Each background is varied individually and the effect on the fitted signal yield is added in quadrature as a contribution to the uncertainty. The higher $\Kstarz$ background is varied by its uncertainty and the largest change to the signal yield is added as a systematic error. The uncertainty due to reconstructing the  wrong $B^0 \rightarrow K^{*0}h^+_1 h^-_2$ signal candidate as a consequence of $K/\pi$ misidentification (for example $B^0 \rightarrow K^{*0}K^+ K^-$ events being reconstructed as $B^0 \rightarrow K^{*0}\pi^+ K^-$) is determined using MC events and added as a systematic uncertainty. We compute uncertainties and corrections to MC using the high statistics calibration channel $B^0 \rightarrow D^-(\rightarrow K^{*0} \pi^-) \pi^+$. Over 7000 events are selected using the $B^0 \rightarrow K^{*0}\pi^+ \pi^-$ selection criteria and requiring the reconstructed $K^{*0}\pi^-$ invariant mass to be in the range $1.84<m_{K^{*0}\pi^-}<1.88 \gevcc$.  The uncertainty due to PDF modeling is estimated from the calibration channel and by varying the PDFs according to the precision of the parameters obtained from the calibration channel fit to data. In order to take correlations between parameters into account, the full correlation matrix is used when varying the parameters. All PDF parameters that are originally fixed in the fit are then varied in turn, and each difference from the nominal fit is combined in quadrature and taken as a systematic contribution.  

\begin{table}[tb]
\caption{Summary of systematic uncertainty contributions to the branching fraction measurements $B^0 \rightarrow K^{*0}h^+_1 h^-_2$. Multiplicative errors are shown as a percentage of the branching fraction and additive errors are shown in events. The final row shows the total systematic error on the branching fraction.}\label{tab:sys}
\begin{center}
\begin{tabular}{ccccc}
\hline
\hline
Error &
\footnotesize{$K^{*0}K^+ K^-$}&
\footnotesize{$K^{*0}\pi^+ K^-$}&
\footnotesize{$K^{*0}K^+ \pi^-$}&
\footnotesize{$K^{*0}\pi^+ \pi^-$}\\
source&
error&
error&
error&
error\\
\hline
\multicolumn{2}{c}{Multiplicative errors (\%)}&
&
 \\
Tracking&
2.0&
2.0&
2.0&
2.0\\
Efficiency&
3.8&
3.9&
3.9&
4.2\\
No. of \BB\ &
1.1&
1.1&
1.1&
1.1\\
\hline
Tot. mult.(\%) &
4.4&
4.5&
4.5&
4.7\\
\hline
\multicolumn{2}{c}{Additive errors (events)}
&
&
 \\
Fit Bias&
14&
10&
2&
43\\
$B$-background&
17&
13&
13&
26\\
Higher $K^{*0}$ bkg&
58&
0&
0&
118\\
Signal mis-id&
0&
6&
11&
0\\
PDF params.&
20&
26&
4&
19\\
\hline
Tot. add.&
\raisebox{-1.5ex}[0cm][0cm]{65}&
\raisebox{-1.5ex}[0cm][0cm]{31}&
\raisebox{-1.5ex}[0cm][0cm]{18}&
\raisebox{-1.5ex}[0cm][0cm]{130}\\
(events) &
&
&
\\
\hline
\hline
Total (10$^{-6}$)&
2.2&
0.8&
0.5&
4.3\\
\hline
\hline
\end{tabular}
\end{center}
\end{table}

Interference effects between the $K^{*0}(892)$ and spin-0 final states (nonresonant and $K^{*0}_0(1430)$) integrate to zero if the acceptance of the detector and analysis is uniform; the same is true of the interference between the $K^{*0}(892)$ and spin-2 final states ($K^{*0}_2(1430)$). Studies of MC events show the efficiency variations are small enough to make these interference effects insignificant.  The integrated interference between $K^{*0}(892)$ and other spin-1 amplitudes such as $K^{*0}(1410)$ is in principle nonzero, but in practice is negligible due to the small branching fraction of $K^{*0}(1410$) $\to K^+\pi^-$ (6.6 $\pm$ 1.3\%~\cite{pdg}) and the fact that the $K\pi$ mass lineshapes have little overlap.

The $C\!P$-violating asymmetries for the decays $B^0 \rightarrow K^{*0}K^+ K^-$, $B^0 \rightarrow K^{*0}\pi^+ K^-$, and $B^0 \rightarrow K^{*0}\pi^+ \pi^-$ are shown in Table~\ref{tab:results} and are consistent with zero. The background asymmetries ${\cal A}^{\mathrm{bkg}}_{K^*KK}$, ${\cal A}^{\mathrm{bkg}}_{K^*K\pi}$, and ${\cal A}^{\mathrm{bkg}}_{K^*\pi\pi}$, which are expected to be consistent with zero, are measured to be 0.017 $\pm$ 0.010, 0.007 $\pm$ 0.004,  and 0.0018 $\pm$ 0.0033, respectively. 

The systematic error on ${\cal A}_{K^*h_1h_2}$ is calculated by considering contributions due to track finding, fit biases, $B$-background uncertainties and particle interaction asymmetries. The error due to fit biases is found to be negligible. Tracking efficiency uncertainties are assigned by comparing the total number of reconstructed tracks for control channels in data and MC simulation. The interaction asymmetry of matter and antimatter with the detector is studied using MC, where biases between -0.01 and -0.03 are observed and applied as corrections to the data. The uncertainty on the correction, obtained from the calibration channel asymmetry difference between MC and data, is added as a systematic uncertainty. The contribution from $B$-background is calculated by varying the number of expected events within errors and by conservatively assuming a large $C\!P$-violating asymmetry of $\pm$ 0.2, as there are no available measurements for these decays.

In summary, we analyze $K^+\pi^- h^+_1 h^-_2$ final states to obtain branching fraction and $C\!P$-asymmetry measurements for the decays $B^0 \rightarrow K^{*0}K^+ K^-$, $B^0 \rightarrow K^{*0}\pi^+ K^-$ and $B^0 \rightarrow K^{*0}\pi^+ \pi^-$. We find the results to be consistent with the Standard Model, observing the decay $B^0 \rightarrow K^{*0}\pi^+ K^-$ with a 5.3 standard deviation significance and placing an upper limit on the branching fraction $B^0 \rightarrow K^{*0}K^+ \pi^-$. We find no evidence for $C\!P$-violation.

We are grateful for the excellent luminosity and machine conditions
provided by our \pep2\ colleagues, 
and for the substantial dedicated effort from
the computing organizations that support \babar.
The collaborating institutions wish to thank 
SLAC for its support and kind hospitality. 
This work is supported by
DOE
and NSF (USA),
NSERC (Canada),
CEA and
CNRS-IN2P3
(France),
BMBF and DFG
(Germany),
INFN (Italy),
FOM (The Netherlands),
NFR (Norway),
MIST (Russia),
MEC (Spain), and
STFC (United Kingdom). 
Individuals have received support from the
Marie Curie EIF (European Union) and
the A.~P.~Sloan Foundation.


\begin{thebibliography}{99}
\bibitem{cc}
The use of charge-conjugate modes is implied throughout this paper unless otherwise noted.
\end{thebibliography}
\end{document}